\documentstyle[epsfig,subfigure,amsmath,amssymb,float]{elsart}
   \begin{document}
   \bibliographystyle{elsart-num}
   \def\bra#1{\mathinner{\langle{#1}|}}
   \def\ket#1{\mathinner{|{#1}\rangle}}
   \def\braket#1{\mathinner{\langle{#1}\rangle}}
   \def\Bra#1{\left<#1\right|} \def\Ket#1{\left|#1\right>}
   {\catcode`\|=\active
   \gdef\Braket#1{\left<\mathcode`\|"8000\let|\BraVert {#1}\right>}}
   \def\BraVert{\egroup\,\mid@vertical\,\bgroup}

   \begin{frontmatter}
   \title{Density profile of a strictly two-dimensional
    Bose gas at finite temperature}
   \author{K. K. Rajagopal}, 
   \author{P. Vignolo}, and
   \author{M. P. Tosi\corauthref{cor1}}
   \corauth[cor1]{Corresponding author, e-mail: {\tt vignolo@sns.it}}
   \address{INFM and Classe di Scienze,\\
    Scuola Normale Superiore,I-56126 Pisa, Italy}
     \date{}
   \maketitle

   \begin{abstract}
   We study a Bose-condensed gas at finite temperature, in which the
   particles of the
   condensate and of the thermal cloud are constrained to move in a
   plane under radial 
   harmonic confinement and interact {\it via} strictly two-dimensional 
   collisions. The coupling parameters are obtained from a calculation 
   of the many-body T-matrix and decreases as temperature increases
   through a dependence 
   on the chemical potential and on the occupancy of excited states.
   We discuss the consequences on the condensate fraction and on the
   density profiles of 
   the condensed and thermal components as functions of temperature,
   within a simplified  
   form of the two-fluid model.

   \end{abstract}
   \end{frontmatter}

   \section{Introduction}

   Recent experiments have realized quasi two-dimensional (2D) cold
   Bose systems 
   by tuning the anisotropy of the trapping potential \cite{gorlitz,safanov} 
   and have stimulated interest in studying Bose-Einstein condensation in
   a low-dimensional regime. An important difference between the macroscopic
   2D fluid and the 3D one is that at finite temperature phase fluctuations 
   in 2D destroy the long-range order, in agreement with the
   Mermin-Wagner theorem 
   \cite{mermin}. A Kosterlitz-Thouless transition to a superfluid state still
   occurs on cooling through the binding of vortices of opposite vorticity
   \cite{kosterlitz} and leads to an algebraic decay of the one-body density
   matrix. However, in a trapped 2D fluid the modifications of the density of
   states caused by the confining potential allow a true condensate to
   exist even 
   at finite temperature.\par

   Another important consequence of lowered dimensionality is that the
   T-matrix for 
   two-body collisions {\it in vacuo} at low momenta and energy, which
   should be used
   to obtain the collisional coupling parameters to lowest order in
   the particle 
   density, vanishes in the strictly 2D limit as the {\it{s}}-wave
   scattering length 
   becomes larger than the width of the axial trapping
   \cite{schick,popov}. It is then 
   necessary to evaluate the scattering processes between pairs of
   Bose particles by 
   taking into account the presence of a condensate and a thermal
   cloud through a  
   many-body T-matrix formalism~\cite{stoof93,bijlsma97,Lee}.\par 
   
   This formalism has already been used in a number of studies of low
   dimensional 
   Bose gases, dealing in particular with phase fluctuations and the
   Kosterlitz-Thouless transition in a variational approach 
   \cite{stoof93,bijlsma97,khawaja,khawaja2}, with a mean-field
   evaluation of the breathing-mode frequency in a trapped 1D gas
   \cite{proukakis}, 
   and with the equilibrium density profile of a 2D condensate
   \cite{Lee,tanatar}. 
   In the limit of zero temperature the condensate-condensate coupling
   parameter has 
   been related to the two-body T matrix by considering that, when two
   particles in the 
   condensate collide at zero momentum, they both require an energy
   equal to the 
   chemical potential  $\mu$ to be excited out of the condensate
   \cite{Lee}. With 
   increasing temperature the population of the excited states becomes
   non-negligible and 
   a microscopic theory which also takes into account the depletion of
   the condensate is required~\cite{proukakis2}. An additional coupling
   parameter to describe the scattering processes between an atom in the
   condensate and an atom in the thermal cloud has been introduced by
   Stoof and co-workers 
   \cite{khawaja} through the two-body T-matrix at energy $-\mu$.\par
   
   The present paper is the finite-temperature extension of the work by
   Tanatar {\it et al.} \cite{tanatar}, who studied the cross-over from a 3D
   to a 2D regime in the equilibrium density profiles at zero temperature. 
   Here we limit ourselves to the strictly 2D case, where the scattering length
   has become larger than the axial thickness of the cloud. We give a
   semi-analytical 
   expression for the many-body T-matrix elements corresponding to the 
   condensate-condensate and to the condensate-thermal cloud coupling
   parameters ($g_{2}$ and  
   $g_{1}$, respectively) and we use them in a two-fluid model
   \cite{minguzzi97} to 
   evaluate the condensate fraction and the density profiles of the
   bosonic gas at 
   increasing temperature. Congruently with the work of Tanatar {\it et al.} 
   \cite{tanatar} we focus on the experimental parameter of G\"orlitz
   {\it et al.} 
   \cite{gorlitz} for the case of largest anisotropy.\par

   The paper is organized as follows. Section. \ref{sec_T} summarizes the
   general treatment of the many-body T-matrix for two-body scattering
   processes 
   in the presence of a condensate and a thermal cloud. In Sec. \ref{sec2} we
   evaluate a simplified form of the two-fluid model for a 2D Bose gas
   at finite 
   temperature. Finally, we present and discuss our results in
   Sec. \ref{conclusion}.  
    
   \section{Coupling parameters in a mixture of condensed and thermal bosons}
   \label{sec_T}
  
   The scattering processes between pairs of atoms in a gas consisting of a 
   Bose-Einstein condensate and a thermal cloud are described by
   the many-body  
   T-matrix $T^{MB}(E)$ as a function of the energy $E$. The coupling
   parameters are given by the matrix elements $\bra{{\bf
   k'}}T^{MB}(E)\ket{{\bf k}} 
   \equiv T^{MB}({\bf k},{\bf k'},{\bf K};E)$, taken in the limit of
   zero energy and  
   momenta. Here ${\bf k}$ and ${\bf k'}$ are the incoming and
   outgoing relative momenta of the
   pair of center-of-mass momentum ${\bf K}$. We shall consider only the
   condensate-condensate and condensate-thermal cloud couplings and neglect
   the scattering between thermally excited atoms in the present case
   of a dilute  
   gas.\par 

   Before discussing the many-body T-matrix, however, we shall first recall the
   behavior of the two-body T-matrix that describes collisions
   between pairs of   
   particles {\it in vacuo}. We shall then discuss how the two
   T-matrices are related 
   in the appropriate limit for a two-fluid system. 

   \subsection{The two-body T matrix}

   The two-body T-matrix is the solution of the Lippmann-Schwinger equation,
   \begin{eqnarray}
    \bra{\bf k'}T^{2B}(\bar{E})\ket{{\bf k}}&=&\bra{{\bf k'}}V(|{\bf
   r}_{1}-{\bf r}_{2}|) 
   \ket{{\bf k}}\nonumber\\
   &+&\sum_{{\bf q}}\bra{{\bf k'}}V(|{\bf r}_{1}-{\bf
   r}_{2}|)\ket{{\bf q}}  
   \frac{1}{\bar{E}-2\epsilon_{\bf q}}\bra{{\bf
   q}}T^{2B}(\bar{E})\ket{{\bf k}}, 
  \label{one}
   \end{eqnarray}
   with $V(|\bf{r}_{1}-\bf{r}_{2}|)$  being the interparticle potential. 
   The center-of-mass energy for the pair of atoms is $\bar{E}$ and
   each atom of mass 
   $m$ has single-particle excitation energy $\epsilon_{{\bf
   q}}=\hbar^2 q^2/2m$, as  
   collision takes place in free space. In Eq. (\ref{one}) the
   collision is described by 
   a single-loop interaction between the two atoms plus contributions
   involving all 
   possible transition routes from state  $\ket{\bf{k}}$ to state
   $\ket{\bf{k'}}$  
   {\it{via}} intermediate states $\ket{\bf{q}}$.\par 
 
   In the case of hard-disk potential of strength $V_{0}=4\pi\hbar^2/m$ and in
   the dilute limit ${\bf k}a,{\bf k'}a\ll 1$  where $a$ is the 2D
   scattering length, 
   the solution of Eq. (\ref{one}) is \cite{khawaja}
   \begin{equation}
    \bra{{\bf k'}}T^{2B}(\bar{E})\ket{{\bf k}}
   \approx \frac{4\pi\hbar^2/m}{\ln|4\hbar^2/\bar{E}ma^{2}|}. 
   \label{two}
   \end{equation}
   In this case the T-matrix is independent of the momenta, but depends on the
   logarithm of the energy and vanishes as $\bar{E}$ approaches zero.
   Therefore, the presence of the surrounding gas must be taken into
   account in  
   the collisional processes. Following the proposal of Morgan {\it{et al.}} 
   \cite{morgan}, this is done by setting $\bar{E}=-2\mu$ for the
   mutual scattering of 
   two condensate particles, this being the energy required for them
   to be excited out 
   of the condensate. By a similar argument Al Khawaja {\it{et al.}}
   \cite{khawaja} set 
   $\bar{E}=-\mu$ for the scattering between a boson in the condensate
   and a boson in 
   the thermal cloud.

   \subsection{The many-body T-matrix}
    
   According to the above argument, a first approximation for the
   many-body T-matrix at 
   zero momenta and energy is
   \begin{equation}
   T_{n}^{MB}(0,0,0;0)=T^{2B}(0,0,0;-n\mu)
   \label{three}
   \end{equation}
   with $n$ = 1 for condensate-thermal cloud scattering and $n$ = 2 for
   condensate-condensate scattering. Further many-body effects can enter the
   scattering processes attended by the presence of a condensate from
   attributing a 
   Bogoliubov spectrum to the intermediate states \cite{stoof93}. At
   low momenta 
   the many-body T-matrix is the solution of the integral equation,
   \begin{eqnarray}
   \bra{\bf{k'}}T^{MB}(E)\ket{\bf{k}}&=&\bra{\bf{k'}}V(|\bf{r}_{1}-\bf{r}_{2}|)
   \ket{\bf{k}}+\sum_{\bf{q}}\bra{\bf{k'}}
    V(|\bf{r}_{1}-\bf{r}_{2}|)\ket{\bf{q}} 
   \nonumber\\ 
   && \times \frac{1+2N(\omega_{\bf{q}})}{E-2\hbar \omega_{\bf{q}}}
   \bra{\bf{q}}T^{MB}(E)\ket{\bf{k}}
   \label{four}
   \end{eqnarray}
   where $\hbar \omega_{{\bf q}}\approx \epsilon_{{\bf q}}+\mu$ are
   the Bogoliubov excitation  
   energies  in the Hatree approximation $\mu\ll\epsilon_{{\bf q}}$
   and $ N(\omega_{{\bf q}}) 
   \approx \{\exp[\beta(\epsilon_{{\bf q}} +\mu)]-1\}^{-1}$ are the
   corresponding population  
   factors. For a contact interaction potential Eq. (\ref{four}) yields
   \begin{equation}
    T^{MB}(0,0,0;E) =\left [ \frac{1}{V_{0}}- \sum_{\bf{q}}
    \frac{1+2N(\omega_{\bf{q}})}{E-2\hbar \omega_{\bf{q}}}\right]^{-1}.
   \label{five}
   \end{equation}
   As discussed by Stoof and Bijlsma \cite{stoof93,bijlsma97},
   Eq. (\ref{five}) is still 
   affected by an ultraviolet divergence, which can be remedied by
   replacing $V_{0}$ in favor 
   of the two-body T-matrix. The final result is
 \begin{equation}
    T^{MB}_{n}(0,0,0;0) =T^{2B}(0,0,0;-n\mu)\left [ 1 +
   T^{2B}(0,0,0;-n\mu)\sum_{\bf{q}} 
    \frac{N(\omega_{\bf{q}})}{\hbar \omega_{\bf{q}}}\right]^{-1},
   \label{six}
   \end{equation}   
  where $T^{2B}(0,0,0;-n\mu)$ is given by Eq. (\ref{two}) with $\bar{E}=-n\mu$.
 
   \subsection{Calculation of coupling parameters}

   The sum in the RHS of Eq. (\ref{six}) can be evaluated analytically
   by replacing the sum 
   over intermediate states by an integral over momentum. Setting
   $N(\omega_{{\bf q}})= 
   \sum_{s=1}^{\infty}\exp(-s\beta\hbar\omega_{\bf{q}})$, we get 
   \begin{equation}
   \sum_{q}\frac{N(\omega_{q})}{\hbar\omega_{q}}= -\frac{m}{2\pi
   \hbar^2}\sum_{s=1}^{\infty}  
    Ei (-s\beta\mu)
   \label{seven}
   \end{equation}
  where $Ei$($x$) is the exponent-integral function. In the asymptotic
   low-temperature regime 
  ($\beta\mu\gg1$) this thermal-population term can be approximated by
   \begin{equation}
   \sum_{s=1}^{\infty} Ei (-s\beta\mu)\rightarrow
     (\beta \mu)^{-1}\ln[1-\exp(-\beta\mu)].
   \label{eight}
   \end{equation} 
   A numerical illustration is given in Fig. 1 for values of the
   system parameters 
   appropriate to the experiments on $^{23}$Na by G\"orlitz {\it et
   al.} \cite{gorlitz}  
   (see Sec. 4). \par
   In summary, the coupling parameters in the 2D Bose gas are given by
    \begin{equation}
    g_{n}^{MB}= \frac{4\pi\hbar^2/m}{\ln|4\hbar^2/(nm\mu
   a^{2})|-2\sum_{s=1}^{\infty} 
    Ei (-s\beta\mu)},
   \label{nine}
   \end{equation}
   with $n$=2 for collisions between pairs of condensate atoms and
   $n$=1 for collisions 
   between an atom in the condensate and a thermally excited
   atom. These parameters depend  
   on temperature both through the chemical potential and through the
   excited-state  
   population factor given in Eq. (\ref{seven}) and asymptotically
   approximated by  
   Eq. (\ref{eight}). If this population factor is dropped, one
   obtains from Eq. (\ref{three}) 
   the corresponding ``two-body'' coupling parameters as 
  \begin{equation}
    g_{n}^{2B}= \frac{4\pi\hbar^2/m}{\ln|4\hbar^2/(nm\mu a^{2})|}\,.
   \label{ten}
   \end{equation}
   In the next Section we evaluate the chemical potential and hence
   the coupling 
   parameters in Eqs. (\ref{nine}) and (\ref{ten}) through a
   self-consistent evaluation  
   of the density profiles in a two fluid-model.

   \section{Equilibrium properties in the two-fluid model}
   \label{sec2}

   Let $n_{c}(r)$ and $n_{nc}(r)$ be the particle density profiles for the
   condensate and for the thermal cloud in a 2D Bose gas which is
   radially confined inside 
   a isotropic planar trap described by the external potential
   $ V_{ext}(r)=m \omega^2 r^2/2$. The two-fluid model
   \cite{minguzzi97,vignolo}  
   combines a solution of the Gross-Pitaevskii equation for the
   condensate with a  
   Hatree-Fock model of the thermal cloud, which is treated as an
   ideal gas subject to  
   an effective  potential $V_{eff}(r)$. In the present case, 
   \begin{equation}
   V_{eff}(r)= V_{ext}(r) + 2g_{1}n_{c} .
   \label{eleven}
   \end{equation} 
    As already noted, we are neglecting the collisions between pairs of bosons
    belonging to the thermal cloud. It is well known that a full
   numerical solution  
   of the Gross Pitaevskii equation can be avoided when the kinetic
   energy term in it 
    can be neglected ~\cite{baym96}. This yields the condensate density in the
    Thomas-Fermi approximation as
   \begin{equation}
   n_{c}(r)= (1/g_{2})[ \mu - V_{ext}(r) - 2g_{1}n_{nc}(r) ]\theta(\mu
   - V_{ext}(r) -  
   2g_{1}n_{nc}(r)).
   \label{twelve}
   \end{equation}
   Since we are interested in examining qualitative behaviors rather
   than in attaining  
   high numerical accuracy, we have adopted Eq. (\ref{twelve}) for the
   condensate density  
   and found that discontinuities occurring in the density profiles at
   the Thomas-Fermi 
   radius can be eliminated by the simple expedient of introducing
   momentum cut-off in  
   the expression for $n_{nc}(r)$. That is, we calculate $n_{nc}(r)$ from   
  \begin{equation}
   n_{nc}(r)=-\frac{m}{2 \pi \hbar^2\beta} \ln \left\{1-
   \exp\left[\beta\left(\mu - V_{eff}(r)- 
         \frac{p_{0}^2}{m}\right)\right]\right\},
   \label{thirteen}
   \end{equation}

  where we take $p_0=\sqrt{2mg_{1}n_{nc}}$ \cite{prokof}. The model is
  then evaluated by solving  
  self-consistently Eqs. (\ref{eleven})-(\ref{thirteen}) together with
  the condition that  
  the areal integral of $n_{c}(r)$+$n_{nc}(r)$ should be equal to the
  total number $N$ of 
  particles.\par
  Before turning to a presentation of our numerical results, let us
  point out the wide 
  range of validity of the present model. A mean-field treatment is
  valid when the 
  diluteness condition  $n_{c}a^2\ll 1$ holds and if the temperature
  of gas is outside 
  the critical region. With regard to the thermal cloud, no
  significant differences have 
  been found between the predictions of the Hatree-Fock and Popov
  approximations in the 
  regime $n_{c}a^2\ll 1$, except at very low temperature where the
  thermal cloud is becoming  
  negligible \cite{dalfovo}.
    
   \section{Results and discussion}
   \label{conclusion}
 
   For a numerical illustration, we have taken the values of particle 
   number, the radial trap frequency, and the scattering length as
   appropriate for 
   $^{23}$Na atoms in the experiment of G\"orlitz {\it et al.}~\cite{gorlitz} 
   ($N =5\times 10^5,\,\omega=188.4\,$Hz, $a=2.8$ nm). Whereas in
   their experiment  
   collisions are in a 3D regime, we focus on a strictly 2D regime
   that could be reached 
   experimentally by increasing either the trap anisotropy parameter
   or the scattering 
   length.\par
   In Fig. 2 we compare the temperature dependence of the coupling parameters 
   $g_{n}^{2B}$ (long dashes for $n$=2 and short dashes for $n$=1)
   with that of the  
   coupling parameters $g_{n}^{MB}$ (full and dotted lines,
   respectively). It is evident 
   that the many-body screening of the interactions due to the
   occupancy of excited 
   states is quite large and rapidly increasing with temperature.\par
   Such many-body screening has, however, very little effect on
   equillibrium properties 
   of the gas for our choice of system parameters. Figure 3 reports
   the condensate 
   fraction $N_{0}/N$ for the $g_{n}^{MB}$ model as a function of
   temperature, in 
   comparison with that of an ideal Bose gas at the same values of the
   system parameters. 
   As is well known, the transition temperature and the condensate
   fraction are lowered by 
   the interactions. However, the $g_{n}^{2B}$ model gives results
   that are practically 
   the same as the $g_{n}^{MB}$ one.\par   
   Finally, the panels in Fig. 4 show the evolution of the density
   profiles 
   for the condensate and for the thermal cloud with increasing
   temperature from near 
   absolute zero to the critical temperature $T_c\simeq 0.95 T_{0}$ 
   with $T_{0}=(\sqrt{6}/\pi)\hbar\omega\sqrt{N}$ being the critical
   temperature of the ideal gas \cite{bagnato}.
   The results in Fig. 4(a) are in good agreement with those of
   Tanatar {\it et al.}~\cite{tanatar}, 
   except that the tails of the profile are missed in the
   Thomas-Fermi approximation. 
   Again the screening of collision from the occupancy of the excited
   states is very 
   small and becomes barely visible at $T \approx$ 0.75 $T_{0}$.\par 
   
   In conclusion, the use of Eq. (\ref{three}) to describe the
   many-body effects in  
   two-body scattering processes in 2D Bose-condensed gas appears to
   be very good 
   in regard to equillibrium properties at large values of particle
   number. A decrease 
   in the number of particles lowers the chemical potential and may
   lead to observable  
   effects for $N\approx$ 10$^{3}$.

   \ack
   This work was partially supported by INFM under the project
   PRA-Photonmatter. Two of us (K.K.R. and M.P.T.) wish to thank the
   Condensed Matter Section of the Abdus Salam International Center
   for Theoretical Physics for their hospitality during the
   preparation of this work.
     
%   \bibliography{tmatrix}

   \newpage

   \centerline{{\bf Figure captions}}

  \noindent {\bf Figure 1.} The correction term
    $A(\beta\mu)$=$\sum_{n=1}^{\infty}Ei(-n\beta\mu)$ from
    excited-state occupancy (dashed line) and its approximate form
    from Eq. (\ref{eight}) (solid line) as functions of $\beta\mu$. In the
    inset a zoom of the region $0.2<\beta\mu<1$ is shown.

   \noindent{\bf Figure 2.} Interaction strengths (in units of
    $\hbar^2/m$) as functions of temperature $T$ (in units of
    ideal-gas critical temperature $T_{0}$).
    
    \noindent{\bf Figure 3.} Condensate fraction $N_{0}/N$ as a
   function of temperature $T$ (in units of $T_{0}$) from the
   $g_{n}^{MB}$ model and (dashed line) compared with the
   non-interacting gas (solid line).

   \noindent{\bf Figure 4.} {Density profile of the condensate (solid
   line) and  
   the thermal cloud (dashed line) in  the $g_{n}^{MB}$ model (in
   units of $N/a^2_{ho}$ 
   with $a_{ho}=(\hbar/m\omega)^{1/2}$) as functions of radial
   distance $r$ (in units 
   of $a_{ho}$) at various values of the temperature
   ($T/T_{0}$=0, 0.3, 0.5, 0.8 and 0.95 
    from (a) to (e)).}

   \newpage
   \begin{figure}
   \centering{
   \epsfig{file=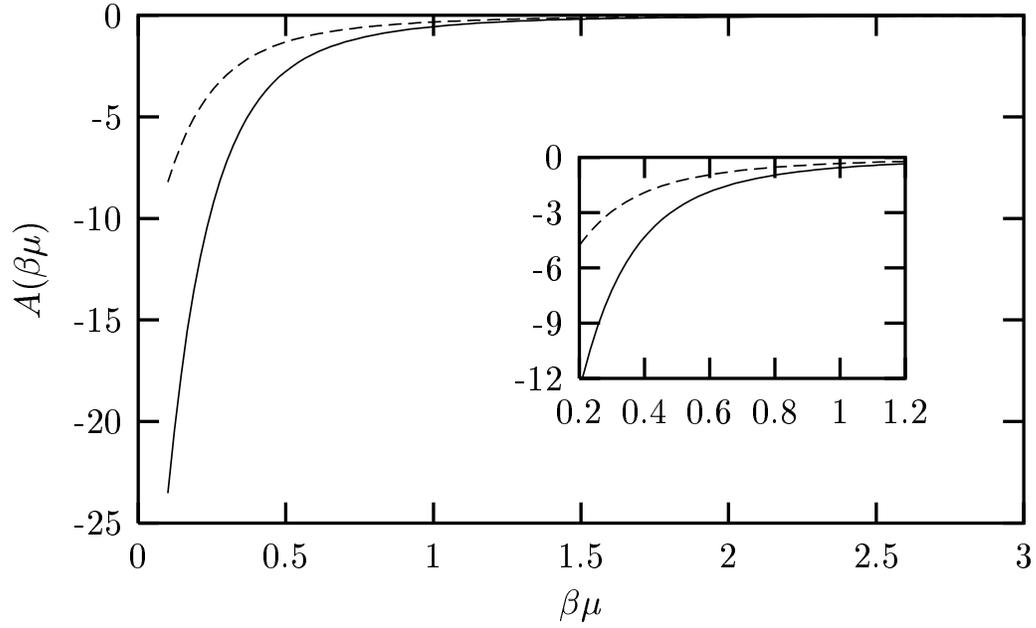,width=1.0\linewidth}}
   \caption{The correction term
   $A(\beta\mu)$=$\sum_{s=1}^{\infty}Ei(-s\beta\mu)$ 
    from excited-state occupancy (dashed line) and its approximate
   form from Eq. (\ref{eight})
    (solid line) as functions of $\beta\mu$. In the inset a zoom of the region
     $0.2<\beta\mu<1.2$ is shown.}
%   \label{fig1}
   \end{figure}

  \begin{figure}
   \centering{
   \epsfig{file=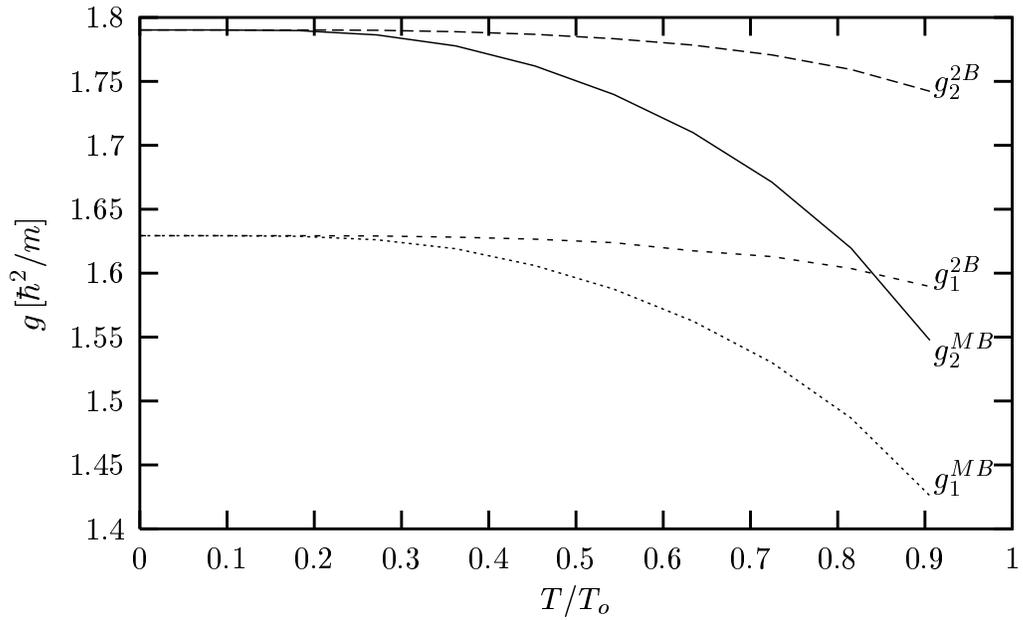,width=1.0\linewidth}}
   \caption{Interaction strengths (in units of $\hbar^2/m$) as
   functions of temperature 
    $T$ (in units of ideal-gas critical temperature $T_{0}$).}
%   \label{fig8}
   \end{figure}

   \begin{figure}
   \centering{
   \epsfig{file=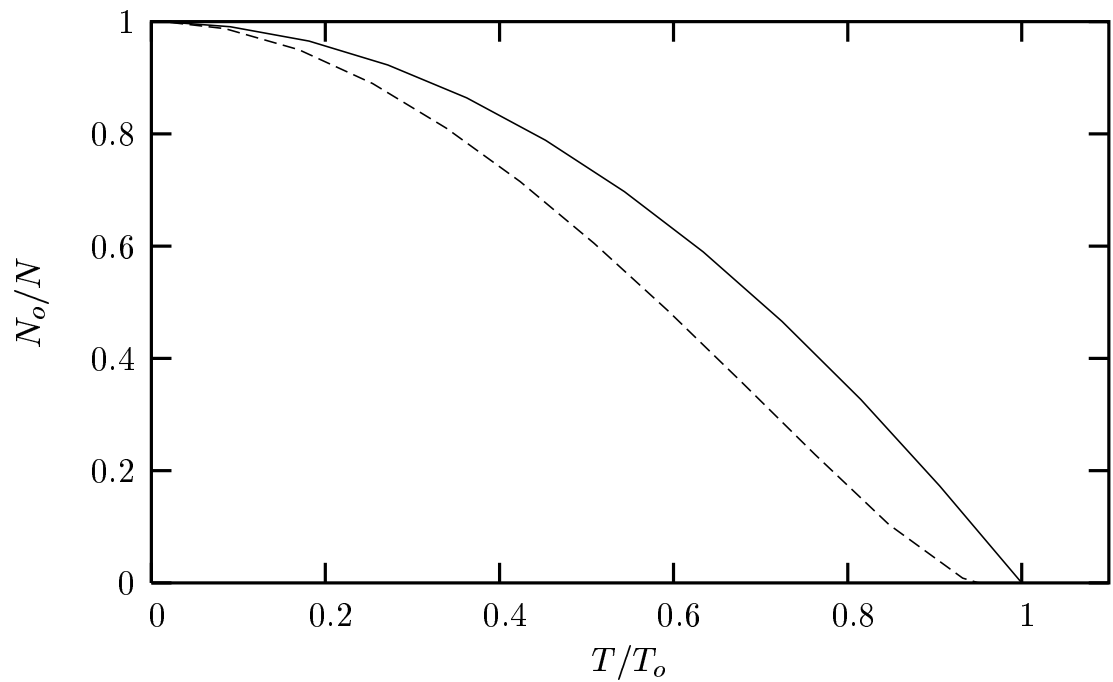,width=1.0\linewidth}}
   \caption{Condensate fraction $N_{0}/N$ as a function of temperature $T$
   (in units of $T_{0}$) from the $g_{n}^{MB}$ model and (dashed line) compared
    with the non-interacting gas (solid line).}
%   \label{fig7}
   \end{figure}

   \begin{figure}
   \centering{
   \subfigure[]{
   \epsfig{file=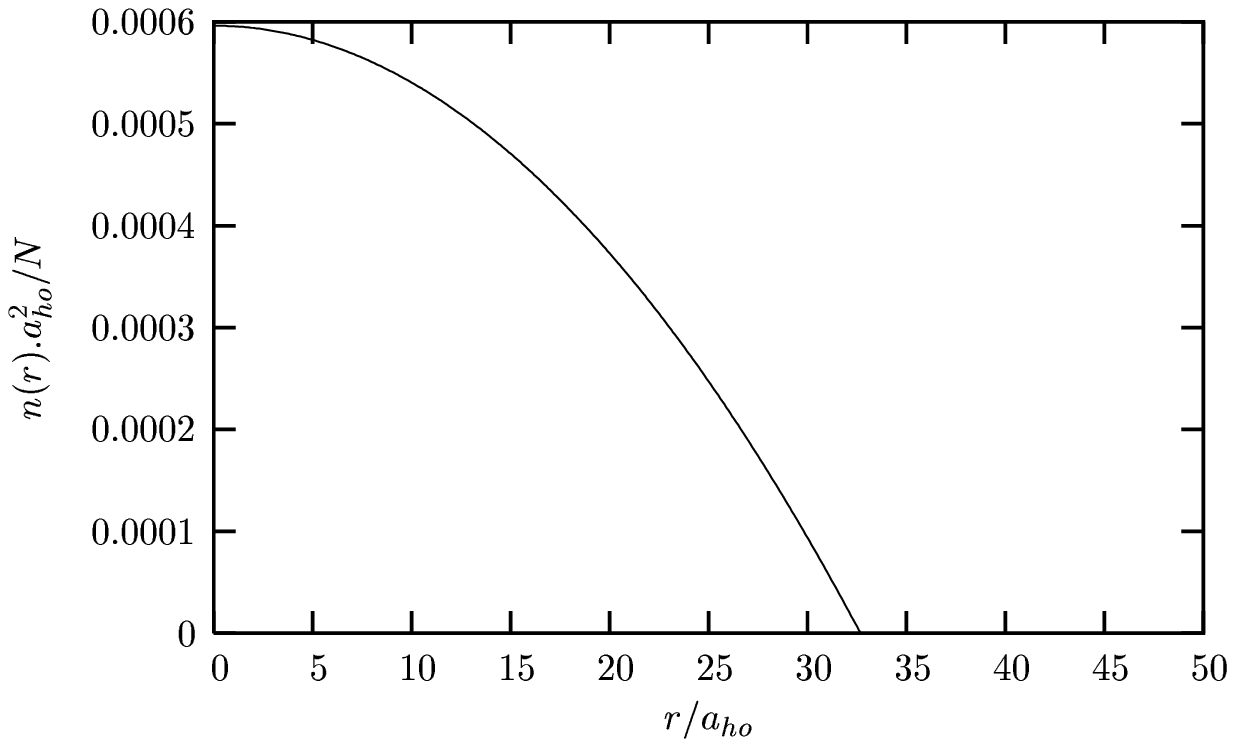,width=0.46\linewidth}}
   \subfigure[]{
   \epsfig{file=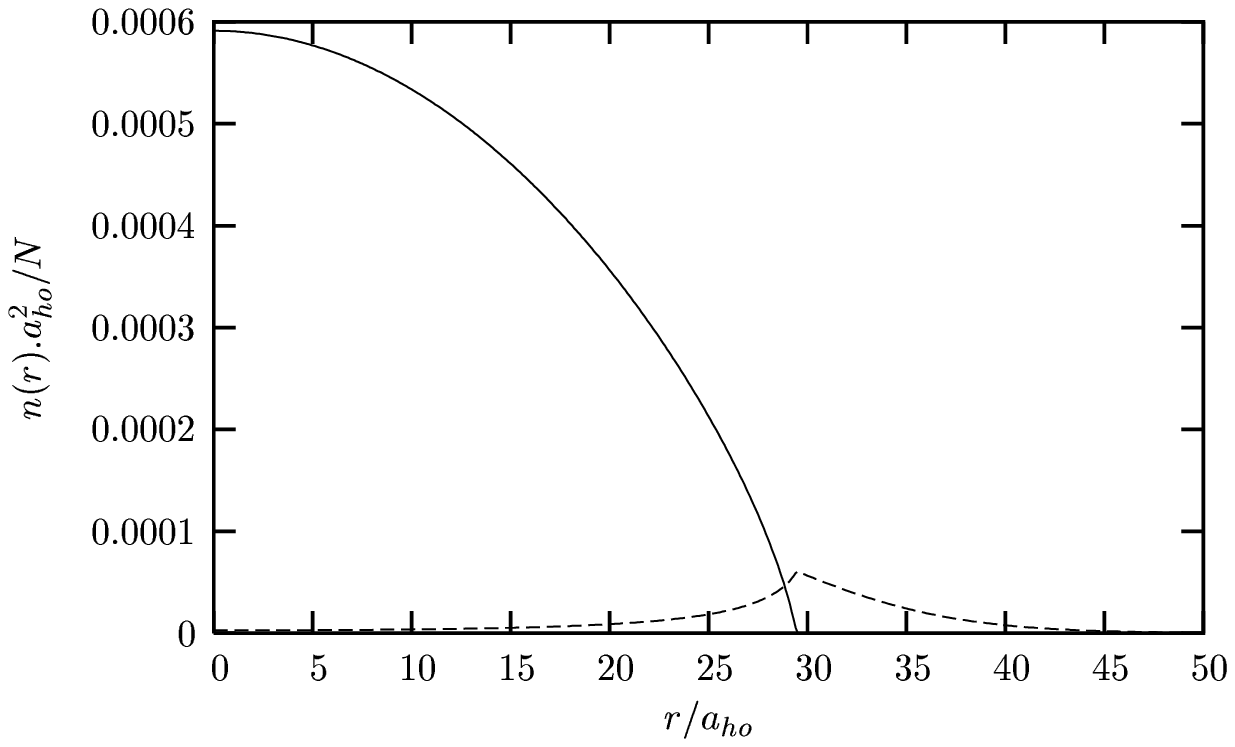,width=0.46\linewidth}}
   \subfigure[]{
   \epsfig{file=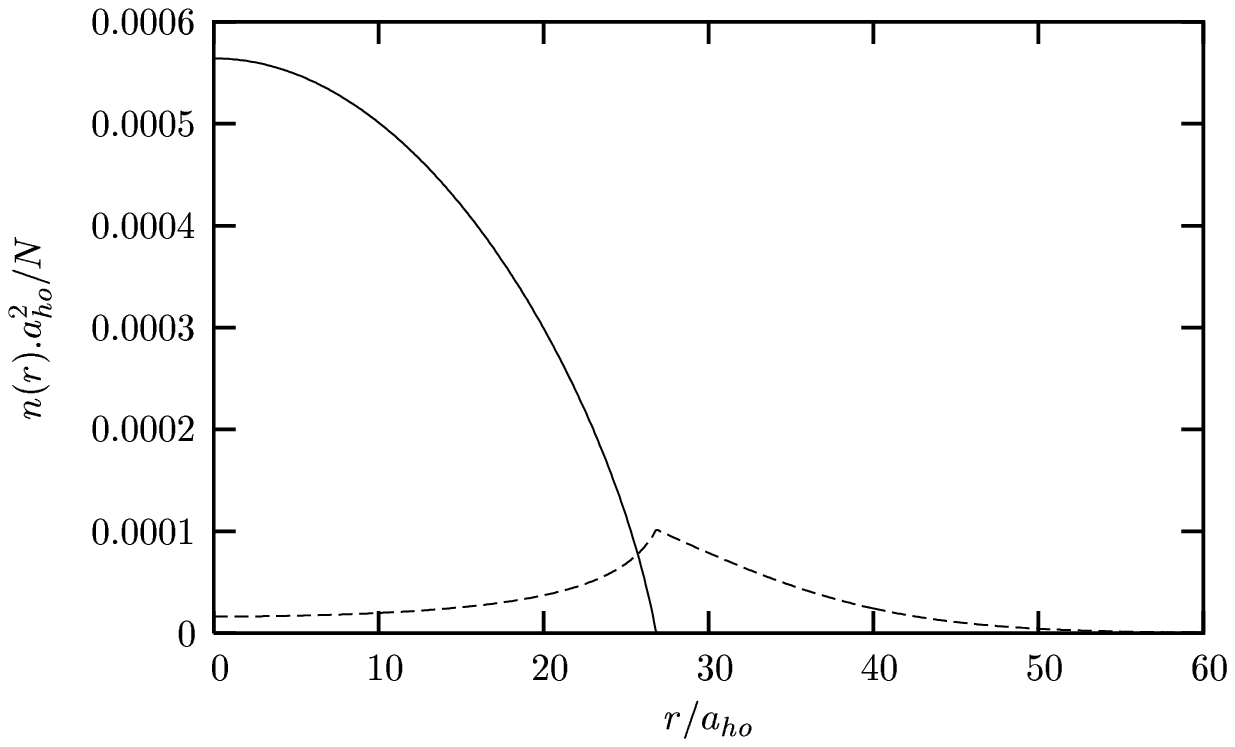,width=0.46\linewidth}}
   \subfigure[]{
   \epsfig{file=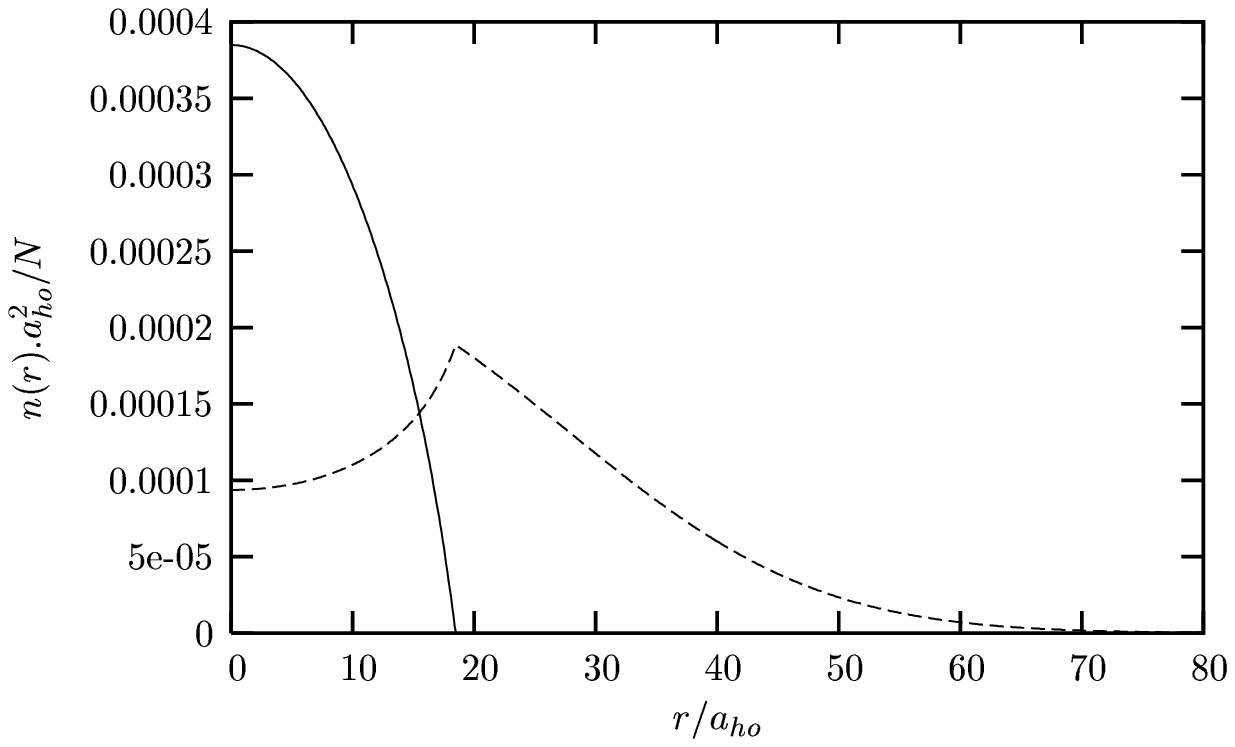,width=0.46\linewidth}}
   \subfigure[]{
   \epsfig{file=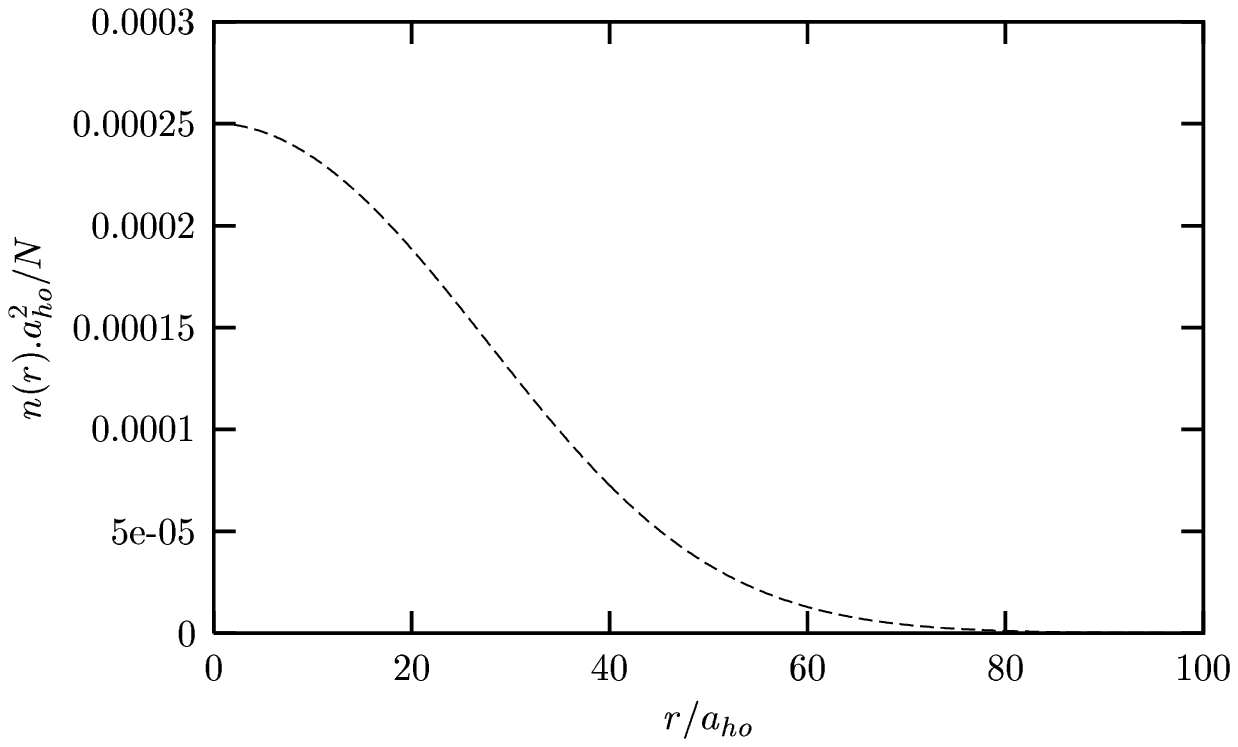,width=0.46\linewidth}}
   }
   \caption{Density profile of the condensate (solid line) and  
   the thermal cloud (dashed line) in  the $g_{n}^{MB}$ model (in
   units of $N/a^2_{ho}$ 
   with $a_{ho}=(\hbar/m\omega)^{1/2}$) as functions of radial
   distance $r$ (in units 
   of $a_{ho}$) at various values of the temperature 
   ($T/T_{0}$=0, 0.3, 0.5, 0.8 and 0.95 
    from (a) to (e)).}
%   \label{fig2}
   \end{figure}

   \end{document}